\documentclass[a4paper]{article}
\usepackage{indentfirst}
\usepackage{fullpage}
\usepackage{amsmath}
\usepackage{amsthm}
\usepackage{latexsym}
\usepackage{epsf}
\usepackage{graphicx,color}

\makeatletter
\renewcommand\paragraph{%
  \@startsection{paragraph}{4}{0mm}%
   {-\baselineskip}%
   {.5\baselineskip}%
   {\normalfont\normalsize\bfseries}}
\makeatother

\begin{document}
%
\newtheorem{lemma}{Lemma}[section]

\newcommand{\bA}{{\bf A}}
\newcommand{\B}{{\bf B}}
\def\bC{{\bf C}}
\newcommand{\bco}{{\boldsymbol{:}}}
\newcommand{\blambda}{{\boldsymbol{\lambda}}}
\newcommand{\bLambda}{{\boldsymbol{\Lambda}}}
\newcommand{\bmu}{{\boldsymbol{\mu}}}
\newcommand{\bn}{{\bf n}}
\newcommand{\bnabla}{{\boldsymbol{\nabla}}}
\newcommand{\bomega}{{\boldsymbol{\omega}}}
\newcommand{\bsigma}{{\boldsymbol{\sigma}}}
\newcommand{\btheta}{{\boldsymbol{\theta}}}
\newcommand{\bo}{{\overline{B(0)}}}
\newcommand{\bu}{{\bf u}}
\newcommand{\bU}{{\bf U}}
\newcommand{\bv}{{\bf v}}
\newcommand{\bw}{{\bf w}}
\newcommand{\bzero}{{\bf 0}}
\newcommand{\ct}{{\mathcal{T}}}
\newcommand{\cth}{{\mathcal{T}_h}}
\newcommand{\dsum}{{\displaystyle\sum}}
\newcommand{\bD}{{\bf D}}
\newcommand{\e}{{\bf e}}
\newcommand{\bF}{{\bf F}}
\newcommand{\bbf}{{\bf f}}
\newcommand{\bG}{{\bf G}}
\newcommand{\bg}{{\bf g}}
\newcommand{\Gx}{{{\overrightarrow{\bf Gx}}}}
\def\hs2{{\hskip -2pt}}

\newcommand{\bI}{{\bf I}} 
\newcommand{\bL}{{\bf L}}
\newcommand{\intbt}{{\displaystyle{\int_{\small B(t)}}}}

\newcommand{\intG}{{\displaystyle{\int_{\Gamma}}}}
\newcommand{\into}{{\displaystyle{\int_{\Omega}}}}
\newcommand{\intobt}{{\displaystyle{\int_{\Omega\setminus\overline{B(t)}}}}}
\newcommand{\intpb}{{\displaystyle{\int_{\partial B}}}}
\newcommand{\lto}{{L^2(\Omega)}}
\newcommand{\no}{{\noindent}}
\newcommand{\obo}{{\Omega \backslash \overline{B(0)}}}
\newcommand{\obt}{{\Omega \backslash \overline{B(t)}}}
\newcommand{\oo}{{\overline{\Omega}}}
\newcommand{\R}{{\text{I\!R}}}
\newcommand{\bs}{{\bf s}}
\newcommand{\bT}{{\bf T}}
\newcommand{\bV}{{\bf V}}
\newcommand{\W}{{\bf W}}
\newcommand{\bx}{{\bf x}}
\newcommand{\bxi}{{\boldsymbol{\xi}}}
\newcommand{\bY}{{\bf Y}}
\newcommand{\bby}{{\bf y}}
\newcommand{\bz}{{\bf z}}
\newcommand{\br}{{\bar{y}}}
\newcommand{\bbu}{{\bar{u}}}

\title{   Dynamics of two disks settling in a two-dimensional narrow channel: 
From periodic motion to vertical chain  in Oldroyd-B fluid}

\author{Tsorng-Whay Pan\footnotemark[1]\\
{\it Department of Mathematics, University of Houston, Houston, Texas 77204, USA} \\ \\
  Roland Glowinski\\
{\it Department of Mathematics, University of Houston, Houston, Texas 77204, USA}\\
{\it Department of Mathematics, Hong Kong Baptist University, Hong Kong}}

\renewcommand{\thefootnote}{\fnsymbol{footnote}}
\footnotetext[1]{pan@math.uh.edu}
\date{}
\maketitle

\begin{abstract}

In this article we present a numerical study of the dynamics of two disks settling in a narrow vertical channel filled  with Oldroyd-B fluid.
Two  kinds of particle dynamics are obtained: (i) periodic interaction between two disks and (ii)  the  chain formation of  two disks. 
For the periodic interaction of two disks,  two different motions are obtained: (a) two disks stay far apart and interact periodically
and (b)   two disks  interact closely and then far apart in a periodic way,  like the drafting, kissing and tumbling of two disks sedimenting 
in Newtonian fluid, due to the lack of strong enough elastic force. For the formation of two disk chain occurred at higher values of the elasticity 
number, it is either a tilted chain or a vertical chain. The tilted chain can be obtained for either that the elasticity number  is less than the 
critical value for having the vertical chain or that the Mach number is greater than the critical value for a long body to fall broadside-on. 
Hence the  values of the elasticity number and the Mach number determine whether the the chain can be formed and the orientation of the chain.

\vskip 1ex


\end{abstract}

\baselineskip 14pt

\setlength{\parindent}{1.5em}

\section{Introduction}

\normalsize
 
The motion of particles in non-Newtonian fluids is not only of fundamental theoretical interest,
but is also of importance in many applications to industrial processes involving particle-laden 
materials (see, e.g., \cite{chhabra1993} and \cite{mckinley2002}).
For example,  during the hydraulic fracturing operation used in oil and gas wells, 
suspensions of solid particles in polymeric solutions are pumped into hydraulically-induced 
fractures. The particles must prop these channels open to enhance the rate of oil
recovery \cite{Economides1989}. During the shut-in stage, proppant settling is pronounced
when the fluid pressure decreases due to the end of hydraulic fracturing process. 
The study of particle chain during settling in vertical channel can help us to understand 
the mechanism of proppant agglomeration in narrow  fracture zones.  There have been   
works on the simulation of the sedimentation of particles in Oldroyd-B fluids in, e.g., \cite{feng1996}, \cite{hu2001}, \cite{huang1998},
\cite{singh2000},  \cite{Yu2004},  \cite{Hao2009} and references given in a review article  \cite{Avino2015}.
Feng et al. study the two-dimensional sedimentation of circular particles in an Oldroyd-B fluid. They obtained chains of two particles aligned with
the direction of sedimentation, which is precisely the microstructure observed in actual experiments  \cite{josliu1994}. In \cite{huang1998}, an arbitrary 
Lagrangian-Eulerian (ALE) moving mesh technique (see \cite{hu2001}) was used to investigate the cross-stream migration and orientations of elliptic 
particles in Oldroyd-B fluids (with and without shear thinning). Huang {\em et al.} found that the orientation of elliptic particles depends 
on two critical numbers, the elasticity and Mach numbers.  In \cite{singh2000},  a fictitious  domain/distributed Lagrange multiplier (FD/DLM) method 
for particulate flow of Oldroyd-B fluids was developed and chains of two particles aligned with the direction of sedimentation were obtained and for the 
case of multiple circular particles, many chains of two particles were found next to the channel walls.  Shao and Yu \cite{Yu2004}  used 
an improved FD/DLM method  to obtain that the stable configuration is the one where the particles are aligned parallel to the flow direction when the Mach number 
the elasticity number are in the range discussed in \cite{huang1998}.

Although numerical methods for simulating particulate flows in Newtonian fluids have been very successful, numerically simulating  particulate flows in viscoelastic
fluids is much more complicated and challenging. One of the difficulties (e.g., see \cite{baaijens1998}, \cite{keunings2000}) for simulating viscoelastic flows is the breakdown
of the numerical methods. It has been widely believed that the lack of positive definiteness preserving property of the conformation tensor at the discrete level during the {\it entire
time integration} is one of the reasons for the breakdown. To preserve the positive definiteness property of the conformation tensor,  several methodologies  have been proposed recently, as 
in \cite{fattal2004}, \cite{fattal2005}, \cite{lee2006} and \cite{lozinski2003}. Lozinski and Owens \cite{lozinski2003} factored the conformation
tensor to get $\bsigma =A A^t$ and then they wrote down the equations for $A$ approximately at  the discrete level. Hence, the positive definiteness of the conformation tensor is forced with 
such an approach. The methodologies developed in   \cite{lozinski2003} have been applied in \cite{Hao2009} together with the FD/DLM method through operator splitting techniques for 
simulating particulate flows in Oldroyd-B fluid.  In this article,  the computational results of two disks settling  in Oldroyd-B fluids are obtained by the numerical method developed in \cite{Hao2009}.

It is known that particle  settling in a viscoelastic fluid can be quite different from that which is observed in Newtonian fluids.  A well-known one for  two disk interaction can be described 
as drafting, kissing and tumbling in Newtonian fluids \cite{fortes1987} and as  drafting, kissing and chaining in viscoelastic fluids.  In this article, we have studied the dynamics of two 
disks settling in a narrow vertical channel filled  with an Oldroyd-B fluid,  which covers not only  the vertical chain of two disks but also  the tilted chain  of two disks and
the periodic motion  of two disks interacting with each other. We have obtained that the formation and orientation of particle chain depends on the value of the elasticity number and 
that of the  Mach number as discussed in  \cite{huang1998} for elliptic particles settling Oldroyd-B fluids since a chain of two disks is a long body even they are loosely coupled. But 
for the lower values of the elasticity number, we have obtained that two disks  interact with each other periodically in two different ways. One of them is that  two disks  
interact far apart in a periodic way, which is close to the one discussed in \cite{Aidun2003}  for two disks settling in Newtonian fluid in the  lower Reynolds number regime  
(not the drafting, kissing and tumbling phenomenon). The other one is close to  the  drafting, kissing and chaining in viscoelastic fluid; but the last part, chaining,  can not be 
completed due to the lack of strong enough elastic force to hold them together. 
The article is organized as follows. In Section 2, we present a FD/DLM formulation for particulate flows in Oldroyd-B fluid and briefly discuss
the associated  the operator splitting technique, the space and time discretization of the FD/DLM formulation. In Section 3, numerical 
results of the cases of sedimentation of  two disks and their chaining are discussed.

\section{Mathematical Formulations and numerical methods}\label{sec2}

\subsection{Governing equations and its FD/DLM Formulation}
\begin{figure}[!tp]
\begin{center}
\leavevmode
\epsfxsize=2.5in
\epsffile{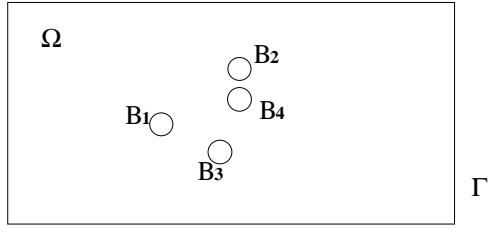}\\
\caption{An example of a two-dimensional flow region with four circular particles.} \label{fig1a}
\end{center}
\end{figure}

Following the work developed  in \cite{Hao2009},  we will first address in the following  the models and computational methodologies.
Let $\Omega$ be a bounded two-dimensional (2D) domain and  let $\Gamma$ be the boundary of $\Omega$.
We suppose that $\Omega$ is filled with a viscoelastic fluid of an Oldroyd-B fluid of density $\rho_f$ and that it contains $N$ moving rigid particles of density
$\rho_s$ (see Figure \ref{fig1a}). 
Let $B(t)=\displaystyle\cup_{i=1}^N B_i(t)$ where $B_i(t)$   is the $i$th
rigid particle in the fluid for $i=1,\dots,N$. We denote by $\partial B_i(t)$  the 
boundary  of  $B_i(t)$ for $i=1,\dots,N$.
For some $T>0,$ the governing equations for the fluid-particle system   are
\begin{eqnarray}
&& \hskip -20pt \rho_{f}( \dfrac{\partial \bu }{\partial t} +
( \bu \cdot \bnabla )\bu ) 
=\rho_{f}\bg -\bnabla p +2\mu \bnabla \cdot \bD(\bu)
+\bnabla \cdot \bsigma^p  \ \  in \ \Omega \backslash \overline{B(t)},\, t\in(0,T), \label{eqn:2.1.1}\\
&& \hskip -20pt \ \nabla \cdot \textbf{u}=0 \ \  in \ \Omega \backslash \overline{B(t)},\, t\in(0,T),\label{eqn:2.1.2}\\
&& \hskip -20pt \ \bu(\bx,0)=\bu_{0}(\bx), \ \  \forall \bx \in \Omega \backslash \overline{B(0)}, \ 
with \, \nabla \cdot \bu_{0}=0, \label{eqn:2.1.3}\\
&& \hskip -20pt \ \bu=\bg_{0} \ \  on \ \Gamma \times
(0, T), with \int _{\Gamma} \bg_{0} \cdot \bn\,
d \Gamma = 0,\label{eqn:2.1.4}\\
&& \hskip -20pt \ \bu= {\bV}_{p,i} + \omega_i \times \stackrel{\longrightarrow}{\bG_i\bx},
 \ \forall \bx \in  \partial B_i(t), \, i=1,\cdots,N,\label{eqn:2.1.5}
\end{eqnarray}
\begin{eqnarray}
&& \hskip -20pt  \dfrac{\partial \bC }{\partial t}+(\bu \cdot \bnabla)\ \bC
-(\bnabla \bu)\ \bC-\bC \ (\bnabla \bu)^t=-\dfrac{1}{\lambda_{1}}(\bC -\bI) \,\,\, in \ \Omega \backslash \overline{B(t)}, \, t\in(0,T),\label{eqn:2.1.6}\\
&& \hskip -20pt \ \bC(\bx,0)=\bC_{0}(\bx),\,  \bx \in \Omega \backslash \overline{B(0)},\label{eqn:2.1.7}\\
&& \hskip -20pt \ \bC=\bC_{L}, \ \ on \ \Gamma^{-},\label{eqn:2.1.8}
\end{eqnarray}
where $\bu$ is the flow velocity, $p$ is the pressure,
$\bg$ is the gravity,   $\mu=\eta_1\lambda_2/\lambda_1$
is the solvent viscosity of the fluid, $\eta=\eta_1-\mu$ is the
elastic viscosity of the fluid, $\eta_1$ is the fluid viscosity,
$\lambda_1$ is the relaxation time of the fluid, $\lambda_2$ is the retardation
time of the fluid, $\bn$ is the outer normal unit vector at $\Gamma,$
$\Gamma^{-}$ is the upstream portion of $\Gamma$.
The  polymeric stress tensor $\bsigma^p$ in 
(\ref{eqn:2.1.1}) is given by $\bsigma^p=\dfrac{\eta}{\lambda_1}  (\bC-\bI)$, where 
the conformation tensor $\bC$ is symmetric and positive definite (see \cite{joseph1990}) 
and $\bI$ is the identity matrix.
 
In (\ref{eqn:2.1.5}), the no-slip condition holds on the boundary  of the $i$th particle,  
$\bV_{p,i}$ is the translation velocity,   $\omega_i$ is the angular velocity and
$\bG_i$ is the center of mass and  $\omega_i \times \stackrel{\longrightarrow}{\bG_i\bx}=(-\omega_i (x_2 - G_{i,2}), \omega_i (x_1 - G_{i,1}))$
(for 2D cases  considered in this article).
The motion of the  particles is modeled by Newton's laws:
\begin{eqnarray} 
&& M_{p,i} \ \dfrac{d \bV_{p,i}}{d t} = M_{p,i}\bg + \bF_i + \bF_i^r,\label{eqn:2.1.9} \\
&&  I_{p,i} \dfrac{d  \omega_i }{d t}= F^t_i,\label{eqn:2.1.10}\\
&& \dfrac{d \bG_i}{d t} = \bV_{p,i},\label{eqn:2.1.11}\\
&& \bG_i(0) = \bG^0_i, \bV_{p,i}(0) = \bV^0_{p,i}, \omega_i(0) = {\omega}^0_i,\label{eqn:2.1.12}
\end{eqnarray} 
for $i=1, \dots, N$,  where in (\ref{eqn:2.1.9})-(\ref{eqn:2.1.12}),  
$M_{p,i}$ and $I_{p,i}$    are the  the mass and the   inertia  of the $i$th particle, 
respectively, $\bF^r_i$ is a short range repulsion  force imposed on the $i$th particle
by other particles and the wall to prevent particle/particle and particle/wall penetration
(see \cite{glowinski2001} for details), and
$\bF_i $ and $ {F}^t_i$ denote the hydrodynamic force and the associated torque
imposed on the $i$th particle by the fluid, respectively.

To avoid the frequent remeshing and the difficulty of the  mesh generation for a time-varying domain 
in which the rigid particles can be very close to each other, especially for 
three dimensional particulate flow, we have extended the governing equations 
to the entire domain $\Omega$ (a fictitious domain). For a fictitious-domain-based variational formulation of 
the governing equations of the particulate flow,  we consider 
only one rigid particle $B(t)$ (a disk in 2D) in the fluid domain without losing generality. 
Let us  define first the following functional spaces
\begin{eqnarray}
&& \bV_{\bg_0(t)}=\{ \ \bv\ |\ \bv\ \in ({H^1(\Omega)})^2, \bv=\bg_0(t) \ on\ \Gamma
\}, \nonumber \\
&& L_0^2(\Omega)=\{ \ q\ |\ q\ \in {L^2(\Omega)}, \int_{\Omega}
q\, d\bx=0 \}, \nonumber \\
&& \bV_{\bC_{L}(t)}=\{ \ \bC\ |\ \bC\ \in ({H^1(\Omega)})^{2 \times 2},
 \bC=\bC_{L}(t) \ on\ \Gamma^{-}\}, \nonumber \\
&& \bV_{\bC_{0}}=\{ \ \bC\ |\ \bC\ \in ({H^1(\Omega)})^{2 \times 2},
 \bC=0 \ on\ \Gamma^{-}\}, \nonumber \\
&& \Lambda(t)={H^1(B(t))}^2. \nonumber
\end{eqnarray}
Following the methodologies developed in \cite{singh2000,glowinski2001}, a fictitious domain 
formulation of the governing equations  (\ref{eqn:2.1.1})-(\ref{eqn:2.1.12}) reads as follows:

{\em For a.e. $t>0,$ find  $\bu(t) \in \bV_{\bg_0(t)}$, $p(t)\in  L_0^2(\Omega)$, $\bC(t)\in
\bV_{\bC_{L}(t)}$, $\bV(t)\in \R^2$, $\bG(t)\in \R^2$, $\omega(t) \in \R$,
 $\blambda(t) \in \Lambda(t)$  such that}
\begin{eqnarray}
&& \begin{cases}
\displaystyle \rho_{f}\int_{\Omega} \left[ \dfrac{\partial \bu}{\partial t} +
(\bu \cdot \bnabla)\bu \right] \cdot \bv\, d\bx 
+2\mu \int_{\Omega}\bD (\bu):\bD (\bv)\,\, d\bx - \int_{\Omega} p \bnabla \cdot \bv\, d\bx \\
\displaystyle -\int_{\Omega} \bv \cdot  (\bnabla \cdot \bsigma^p )\,\, d\bx +
 (1-\rho_f/\rho_s) \lbrace M_p \dfrac{d\bV}{dt} \cdot \bY 
 + I_p \dfrac{d\omega}{dt}\cdot \theta \rbrace \\
\displaystyle -<\blambda, \bv-\bY -\theta \times {\stackrel{\longrightarrow}{\bG\bx}} >_{B(t)} 
 -\bF^r \cdot \bY \\
\displaystyle =\rho_{f} \int_{\Omega} \mathbf{g} \cdot \mathbf{v}d \mathbf{x} 
+(1-\rho_f/\rho_s)M_p \bg \cdot \bY,\\
\displaystyle \forall \{\bv, \bY, \theta\} \in 
(H_0^1(\Omega))^2 \times \R^2 \times \R,
\end{cases}  \label{eqn:2.1.24}\\
&& \int_{\Omega} q \bnabla \cdot \bu(t) \, d \bx = 0, 
 \forall  q \in L^2(\Omega),  \label{eqn:2.1.25} \\ 
&& <\bmu, \bu(\bx, t) - \bV(t) - \omega(t) \times {\stackrel{\longrightarrow}{\bG(t)\bx}} >_{B(t)}=0,  \ \ \forall \bmu  \in  \bLambda(t),  \label{eqn:2.1.26}\\
&&\int_{\Omega} \left( \dfrac{\partial \bC }{\partial t}+(\bu \cdot \bnabla)\bC 
-(\bnabla \bu)\bC-\bC (\bnabla \bu)^t \right) 
:\bs \, d\bx \label{eqn:2.1.27}\\
&& = -\int_{\Omega} \dfrac{f(\bC)}{\lambda_{1}}(\bC- \bI):\bs \, d\bx,
\forall \bs \in \bV_{\bC_0}, \ with\ \ \bC=\bI \ \ in\ \ B(t), \nonumber \\
&& \dfrac{d \bG}{dt} =\bV,   \label{eqn:2.1.28} \\
&& \bC(\bx, 0) = \bC_0(\bx), \forall \bx \in \Omega,  \ with\ \ \bC_0=\bI \ \ in\ \ B(0),  \label{eqn:2.1.29}\\ 
&& \bG(0) = \bG_0, \ \bV(0) = \bV_0, \ \omega(0) = \omega_0, \ B(0) = B_0, \label{eqn:2.1.30}\\
&& \bu(\bx, 0) = \begin{cases}\bu_0(\bx), \forall \bx \in \Omega\setminus{\overline {B_0}}, \\
 \bV_0 + \omega_0 \times {\stackrel{\longrightarrow}{\bG_0\bx}}, \ \forall \bx \in {\overline{B_0}}.
\end{cases} \label{eqn:2.1.31}
\end{eqnarray}
In (\ref{eqn:2.1.24}) the {\it Lagrange multiplier} $\blambda$  defined over $B$ can be viewed as
an extra body force maintaining the rigid body motion inside $B$. The conformation tensor
$\bC$ inside the rigid particle is extended as the identity tensor $\bI$
as in (\ref{eqn:2.1.27}) since the polymeric stress tensor is zero inside the rigid particle.
In equation (\ref{eqn:2.1.24}), since $\bu$ is divergence free and satisfies the Dirichlet 
boundary conditions on $\Gamma,$ we have 
$2 \int_{\Omega}\bD (\bu):\bD (\bv)
\, d\bx =\int_{\Omega}  \bnabla \bu: \bnabla \bv \, d\bx, \, \forall \bv \in (H_0^1(\Omega))^2$.
This is a substantial simplification from the computational point of view, which is another
advantage of the fictitious domain approach. With this simplification, we can use fast solvers for the elliptic problems in order to speed up computations.
Also the gravity term $\bg$ in ($\ref{eqn:2.1.24}$) can be absorbed in the pressure term. 
 
The details of numerical methodologies for simulating the motion of disks sedimenting in Oldroyd-B fluid
in a vertical two-dimensional channel are given in \cite{Hao2009}.  Applying Lie's scheme to 
(\ref{eqn:2.1.24})-(\ref{eqn:2.1.31}),  we have used a  seven stage operator-splitting scheme to obtain numerical results, namely:  
 In Stage 1, we use a Neumann preconditioned Uzawa/conjugate gradient algorithm to
force (in a $L^2$  sense)  the incompressibility condition of $\bu$ discussed in \cite{Hao2009} and \cite{glowinski2003}. 
 In Stage 2, we combine two advection steps: one for $\bu$ and one for $\bC$, which are solved by a wave-like equation method
(see  \cite{glowinski2003} and   \cite{dean1997})  which is an explicit method and does not introduce numerical dissipation.  In this stage, we have transformed 
the advection step for $\bC$ into the one for $A$ like Lozinski and Owens' work in  \cite{lozinski2003}  where  $\bC$ is  factorized as $A A^t$.
In Stage 3, we combine a diffusion step for $\bu$ with a step taking into account the remaining operator 
in the transformed evolution equation verified by $A$. 
In Stage 4, we update the position of $\bG$. 
In Stage 5, we force the rigid body motion of the particle and update $\bV$ and $\omega$ by a conjugate gradient method 
given in  e.g., \cite{Hao2009} and \cite{glowinski2003}, and then impose the condition $\bC = \bI$ inside the particle. 
In Stage 6, we correct the position of $\bG$ via the updated $\bV$ and $\omega$.
Finally,  Stage 7 is a diffusion step for the velocity, driven by the updated polymeric stress tensor.

  \begin{figure}[tp]
     \begin{center} 
     \includegraphics[width=0.125\textwidth]{2disk-case1e0d10-1.eps}
     \includegraphics[width=0.125\textwidth]{2disk-case1e0d10-2.eps}
     \includegraphics[width=0.125\textwidth]{2disk-case1e0d10-3.eps}
     \includegraphics[width=0.125\textwidth]{2disk-case1e0d10-4.eps}     
     \includegraphics[width=0.125\textwidth]{2disk-case1e0d10.eps}       
     \end{center}
     \vskip -2ex
     \caption{Positions of two disks (left four) and trajectories of two disks  (right) for $\rho_s=1.0025$ and  E=0.16 (the other associate  numbers are Re=0.3858, M=0.1453 and De=0.0617).}    \label{fig2a}                                     
     \begin{center} 
     \includegraphics[width=0.125\textwidth]{2disk-case1e0d16-1.eps}
     \includegraphics[width=0.125\textwidth]{2disk-case1e0d16-2.eps}
     \includegraphics[width=0.125\textwidth]{2disk-case1e0d16-3.eps}
     \includegraphics[width=0.125\textwidth]{2disk-case1e0d16-4.eps}     
     \includegraphics[width=0.125\textwidth]{2disk-case1e0d16.eps}       
     \end{center}  
     \vskip -2ex
     \caption{Positions of two disks (left four) and trajectories of two disks  (right) for$\rho_s=1.0025$ and  E=0.256 (the other associate  numbers are Re=0.7222, M=0.3654 and De=0.1848).}    \label{fig2b}
\end{figure}     
 
 \begin{figure}[tp]
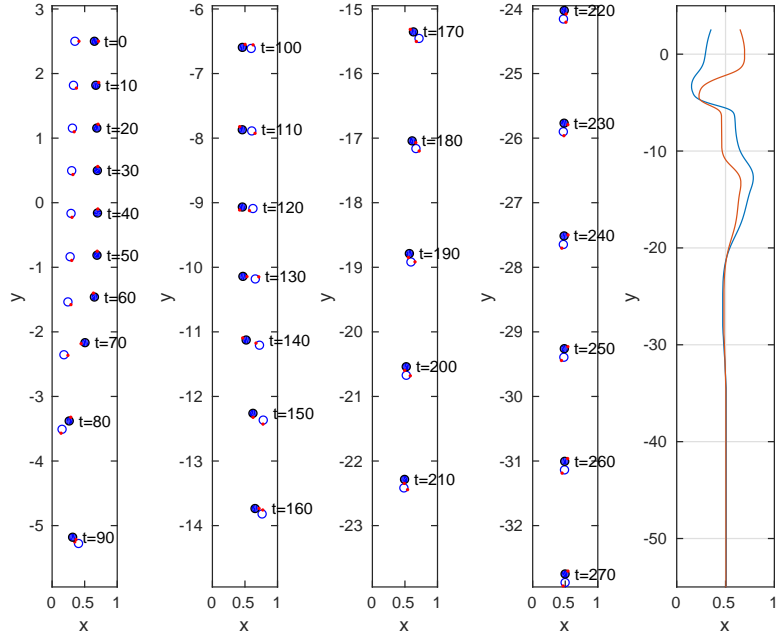

     \begin{center} 
     \includegraphics[width=0.125\textwidth]{2disk-case1e0d20-1.eps}
     \includegraphics[width=0.125\textwidth]{2disk-case1e0d20-2.eps}
     \includegraphics[width=0.125\textwidth]{2disk-case1e0d20-3.eps}
     \includegraphics[width=0.125\textwidth]{2disk-case1e0d20-4.eps}     
     \includegraphics[width=0.125\textwidth]{2disk-case1e0d20.eps}       
     \end{center}  
     \vskip -2ex
     \caption{Positions of two disks (left four) and trajectories of two disks  (right) for$\rho_s=1.0025$ and  E=0.32 (the other associate  numbers are Re=0.8731, M=0.4939 and De=0.2794).}    \label{fig2c}
\end{figure}     
 
  \begin{figure}[tp]
     \begin{center} 
     \includegraphics[width=0.32\textwidth]{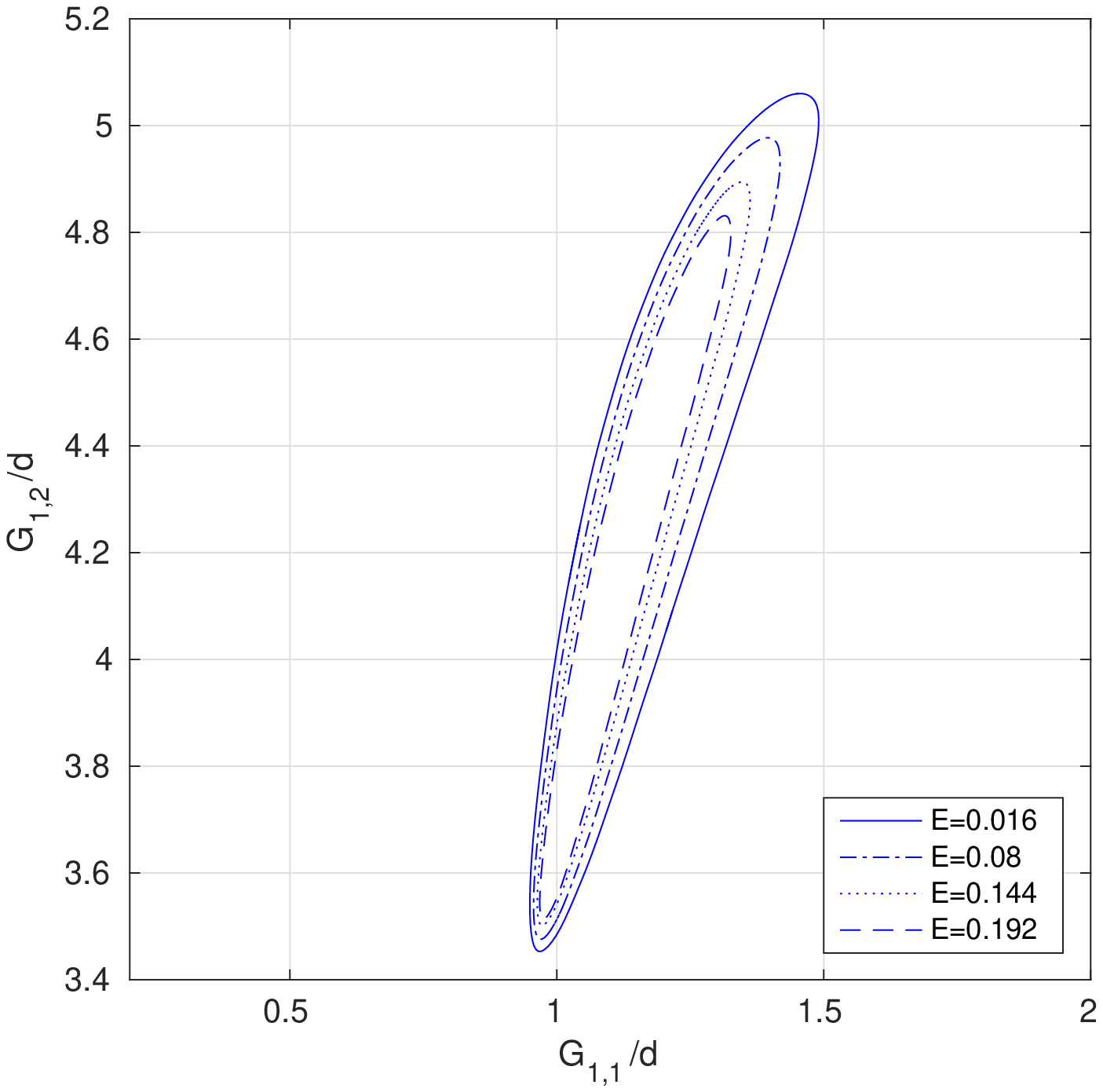}    \hskip 10pt 
     \includegraphics[width=0.32\textwidth]{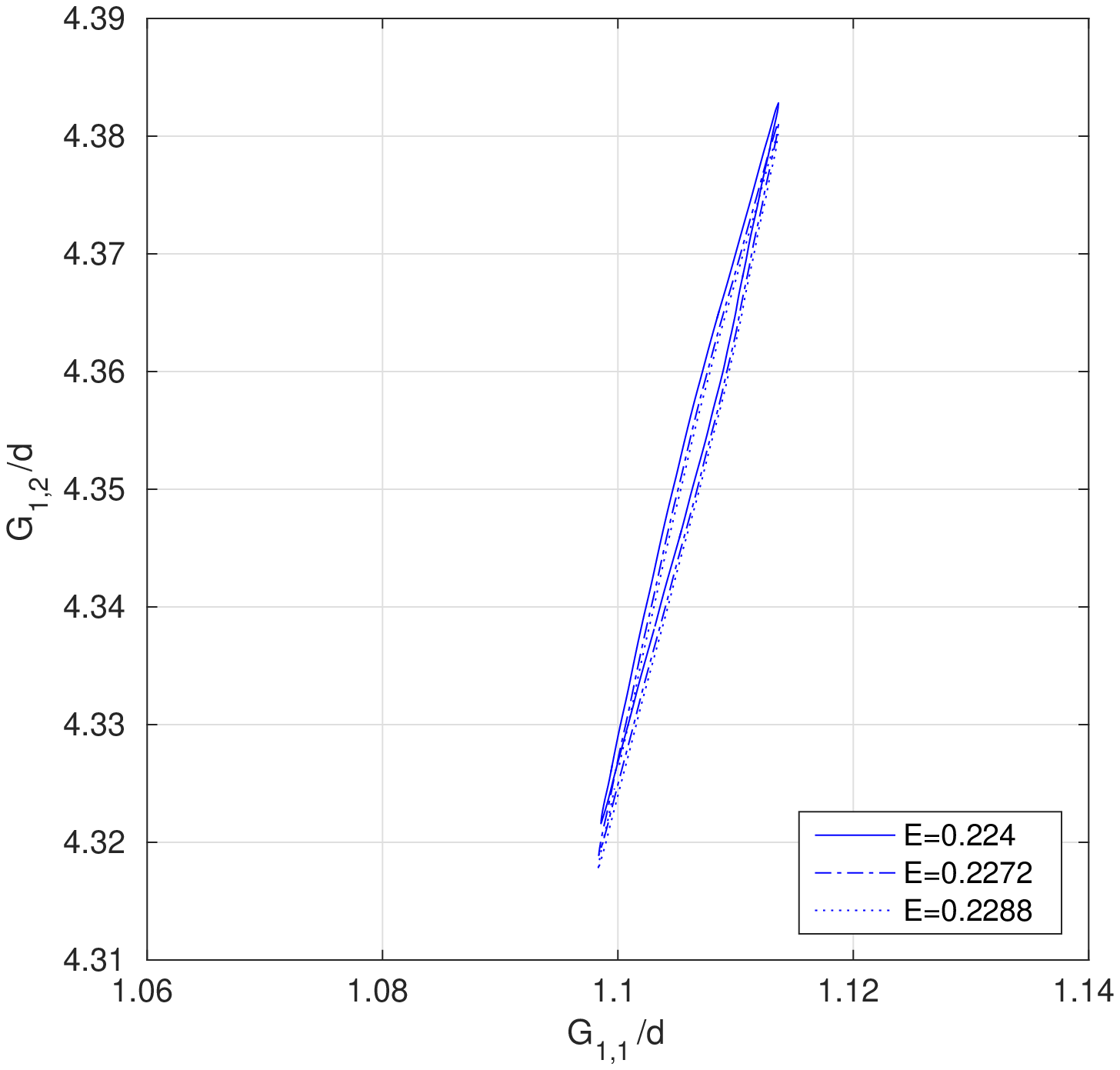}

    \end{center}  
     \caption{Trajectories of disks in phase space of the normalized distances  between the disk mass center and the left wall 
                   for $\rho_s=1.0025$.  The periods are 58.55, 56.80, 55.485 and 54.205 for  E=0.016, 0.080, 0.144 and 0.192, 
                   respectively (left) and  the periods are 52.067, 49 and 47.9 for E=0.224, 0.2272 and  0.2288, respectively (right).}    \label{fig2d}
\end{figure}

\section{Numerical Results and discussion}

In the following discussion, the  particle Reynolds number is Re$=\dfrac{\rho_f U d}{\eta_1}$, the Debra number is 
De$=\dfrac{\lambda_1 U}{d}$, the Mack number is M=$\sqrt{\rm De Re}$, and the elasticity number is 
E=De/Re=$\dfrac{\lambda_1 \eta_1}{ d^2 \rho_f}$  where $U$ is  the averaged terminal speed of disks and $d$ is the disk diameter. 
As discussed in \cite{huang1998} and \cite{Liu1993},  when the elasticity number E  is larger than the critical value ($O(1)$), a long body 
settling in Oldroyd-B fluids  turns its broadside parallel to the flow  direction.   But for the elasticity number E less than the critical value,
it falls steadily in a configuration in which the axis of the long body is at a fixed angle of tilt with the horizontal.  Also for  larger Mach numbers, 
the long body flips into broadside on falling again.  For the dynamics of two disks  settling in Oldroyd-B fluid, they can be viewed as a long body 
if they form a chain.  We like to study the equilibrium orientation of this ``long body'' of two disk chain by varying the elasticity number.  But two disks can 
stay separated for the smaller value of E, it is interesting to find out how two disks interact and whether such interaction is close to the dynamics of two settling disks 
in Newtonian fluid such as drafting, kissing and tumbling phenomenon  \cite{fortes1987} or some kinds of periodic motion discussed in \cite{Aidun2003}.
The computational results of two disk motion in this section are obtained by the numerical method developed in \cite{Hao2009}.

For all numerical results, we have   considered the  settling of two disks in a vertical channel of  infinite length filled with an Oldroyd-B fluid as in \cite{Hao2009}, 
the computational domain is $\Omega= (0,1)\times(0,6)$ initially and then it moves vertically with the mass center of the disk  (see, e.g., \cite{Hu1992} 
and \cite{Pan2002} and references therein for adjusting the  computational domain according to the  position of the particle).   The two disk diameters 
are $d=$0.125 and the initial position of the disk centers are at (0.35, 2.5) and (0.65, 2.5), respectively.  The disk density $\rho_s$ is 1.0025 for first 
several cases considered in this section and the fluid density $\rho_f$ is 1. The fluid viscosity $\eta_1$ is 0.025.  The relaxation time $\lambda_1$ varies 
between 0.01 and 1.0 and the retardation time $\lambda_2$ is $\lambda_1/4$. Then the  associated elasticity number is  E=$1.6 \lambda_1$.  
In Figs. \ref{fig2a},  \ref{fig2b} and \ref{fig2c},  three typical motions of two disks settling in Oldroyd-B fluid are presented. 
For E=0.16 ($\lambda_1=0.1$),  the dynamics of  two disk interaction is characterized by its periodical motion (of period 55.25 time units) as in 
Fig. \ref{fig2a}, which is similar to the one of the motions of two settling disks  in a Newtonian fluid obtained in \cite{Aidun2003}.    
For a slightly higher value, E=0.256 ($\lambda_1=0.16$), two disks form a chain with a stable  tilt angle of 29.39 degrees (see Fig. \ref{fig2b}),  
which is similar to the behavior of a long body when the elasticity number is less than the critical value for turning  its broadside parallel to the flow  direction. 
For E=0.32 ($\lambda_1=0.2$),  we have obtained that two disks form a stable vertical chain as shown in Fig. \ref{fig2c}, which indicates the 
 the critical value of the elasticity number for having a vertical chain  is somewhere between 0.256 and 0.32.
 
For  more details  about the dynamics of two disks, we  have obtained the following results by  varying the relaxation time $\lambda_1$ (resp., the 
elasticity number) from 0.01  to 1  (resp., from 0.016  to 1.6). For E   between 0.016 ($\lambda_1$=0.01) and 0.24 ($\lambda_1$=0.15), two disks stay separated 
and their  interaction is periodical. For example, in the phase space shown in Fig. \ref{fig2d}, constructed by the normalized distances  between the disk mass 
center ($\bG_i=\{G_{1,i},G_{2,i}\}^t$, $i=$1, 2)  and the side wall, i.e., $G_{1,1}/d$ and $G_{1,2}/d$, the attractor  is a limit cycle for each value 
of the elasticity number. At E=0.208 ($\lambda_1$=0.13), the limit cycle shrinks to about a point. Then another kind of   limit cycle occurs  for  
$0.208 < {\rm E} \le 0.2288$ (see Fig.  \ref{fig2d}). But for $0.2304 \le {\rm E} \le 0.24$, two disks settle without noticeable periodic motion
and remain separated at a constant distance. The gap between two disk decreases when increasing the value of E  from  0.2304 to 0.24.
For  E=0.256 ($\lambda_1$ = 0.16) and  0.288 ($\lambda_1$ =0.18),  two disks form a chain with a stable  tilt angle of 29.69 and 82.33 degrees, respectively  
(see Fig. \ref{fig2b} for  E=0.256). Finally, for  E between  0.304 ($\lambda_1$=0.19 ) and 1.6 ($\lambda_1$=1), a vertical chain of two disks is formed. 
Thus the critical value of the elasticity number for having a vertical chain is somewhere in  the interval [0.288, 0.304].

    \begin{figure}[tp]
     \begin{center} 
      \includegraphics[width=0.48\textwidth]{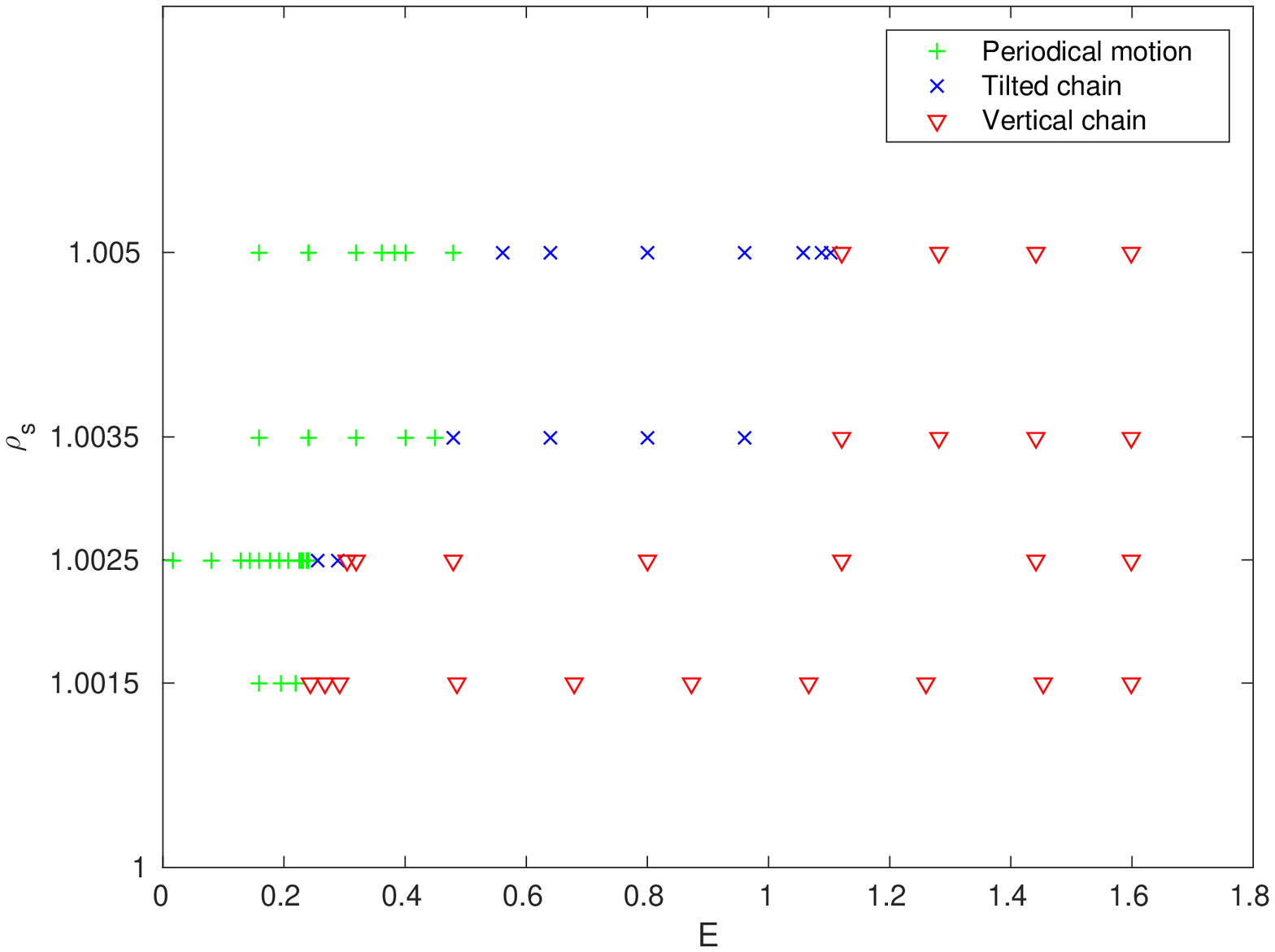}    \hskip 10pt 
       \includegraphics[width=0.48\textwidth]{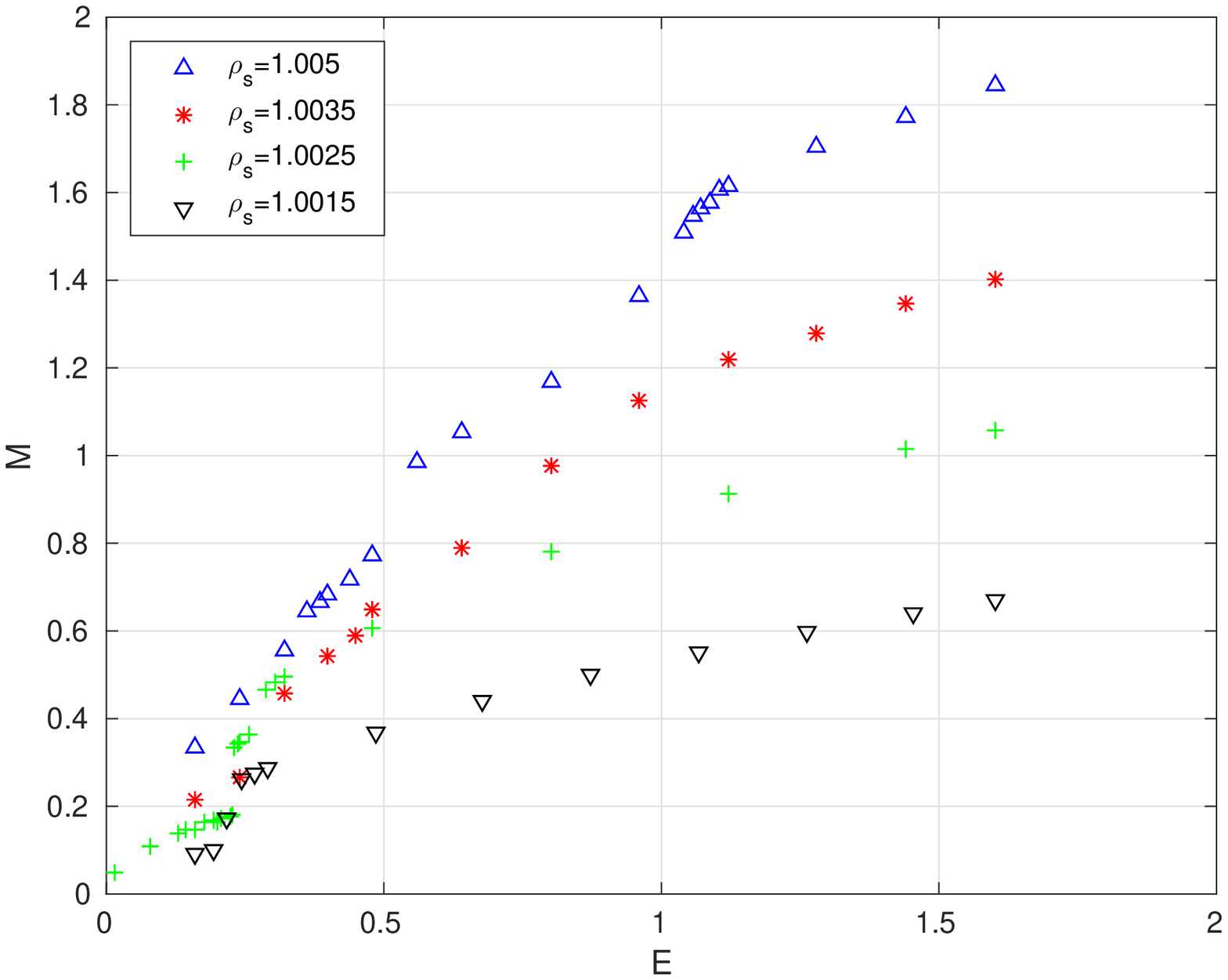} 
   \end{center}  
     \caption{The phase diagram (left) and the associated values of the Mach number (right) of two disk interaction  in a narrow vertical channel for $\rho_s=1.0015$, 1.0025, 1.0035 and 1.005.}    \label{fig2e}
\end{figure}

 \begin{figure}[tp]
     \begin{center} 
     \includegraphics[width=0.15\textwidth]{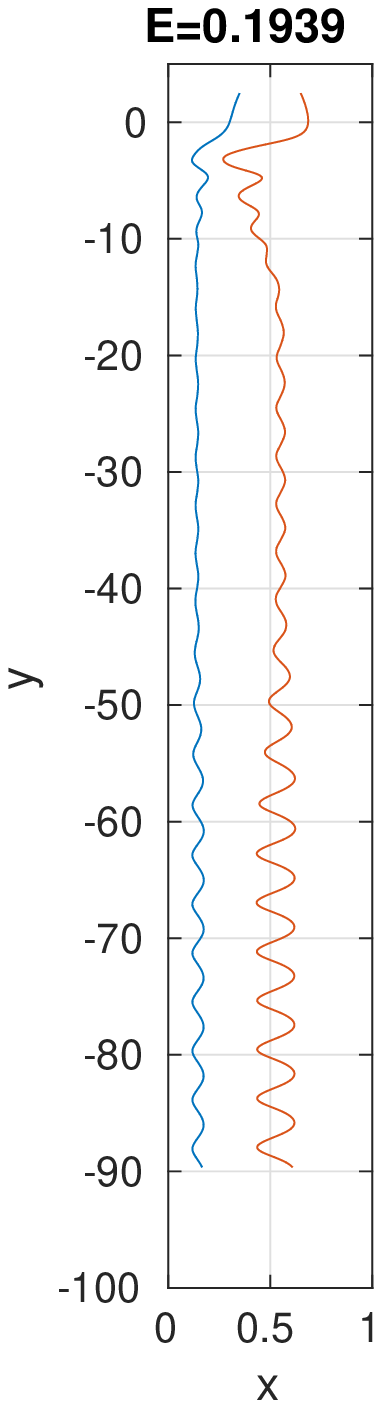}
     \includegraphics[width=0.15\textwidth]{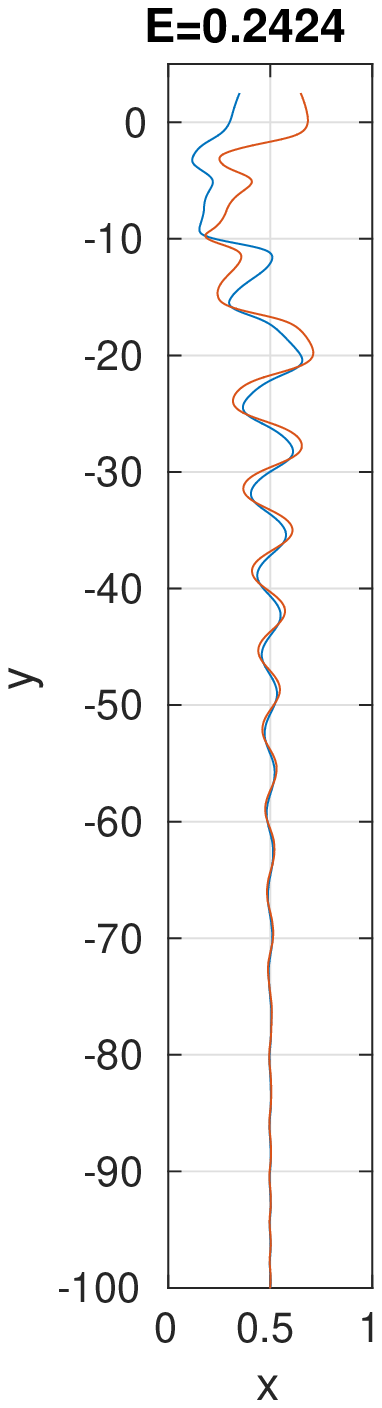}
     \includegraphics[width=0.15\textwidth]{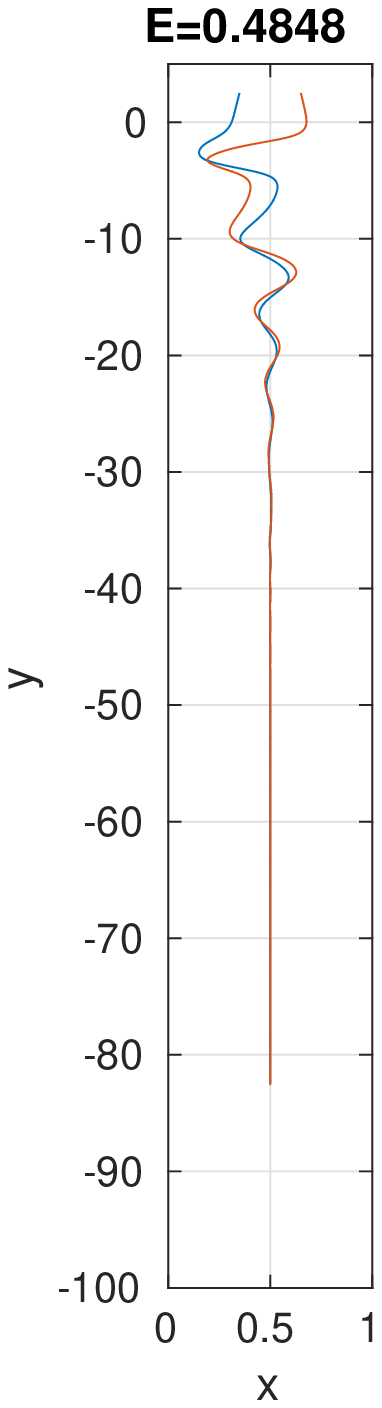}     
     \includegraphics[width=0.15\textwidth]{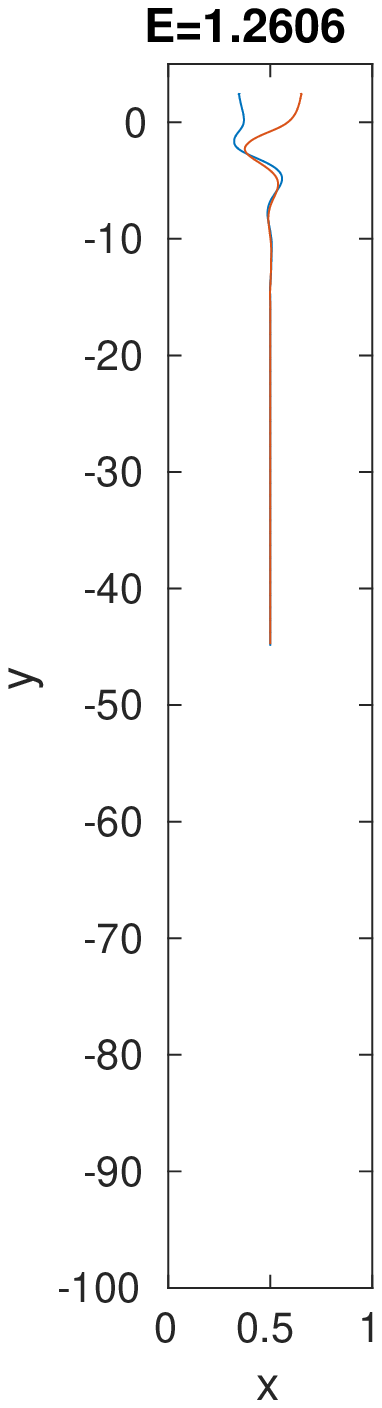}       
     \end{center}  
     \vskip -2ex
     \caption{Trajectories of two disks   for $\rho_s=1.0015$ and different values of E.}    \label{fig2f}
\end{figure}     

 \begin{figure}[!tp]
     \begin{center} 
     \includegraphics[width=0.125\textwidth]{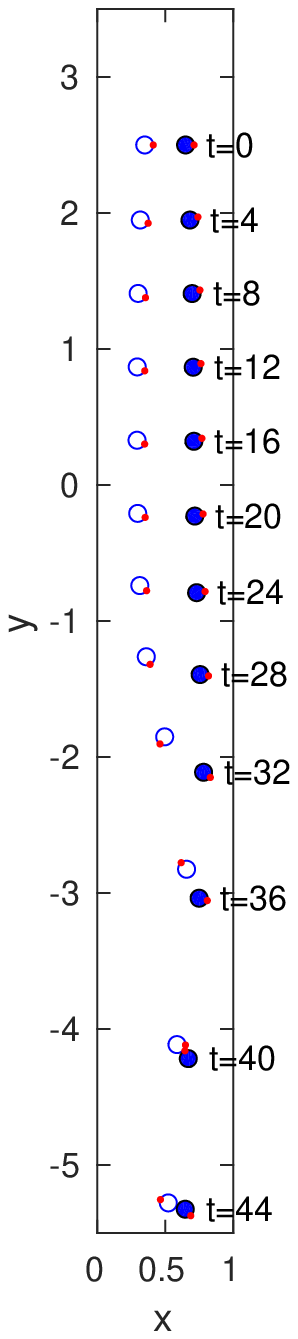}
     \includegraphics[width=0.125\textwidth]{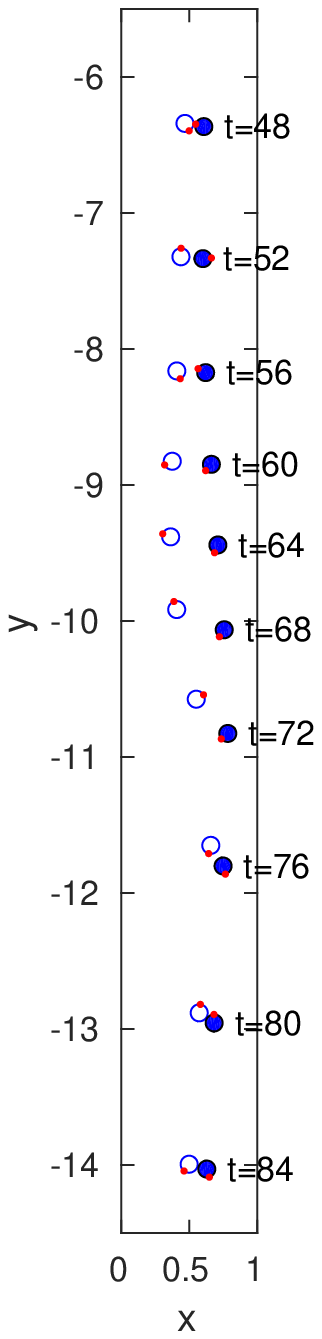}
     \includegraphics[width=0.125\textwidth]{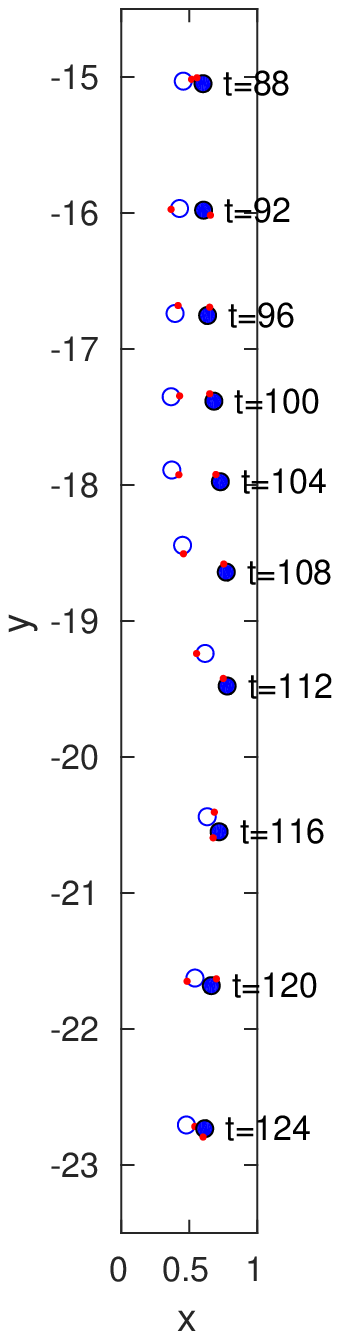}
     \includegraphics[width=0.125\textwidth]{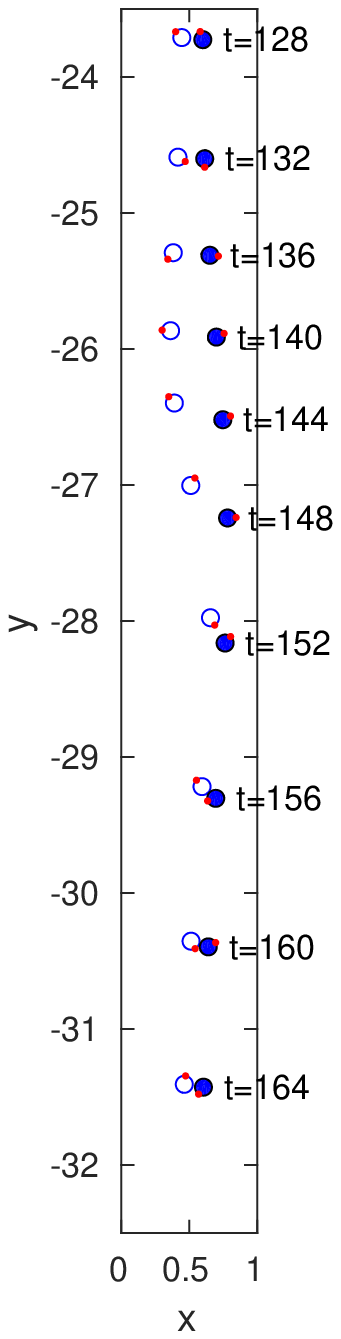}     
     \includegraphics[width=0.135\textwidth]{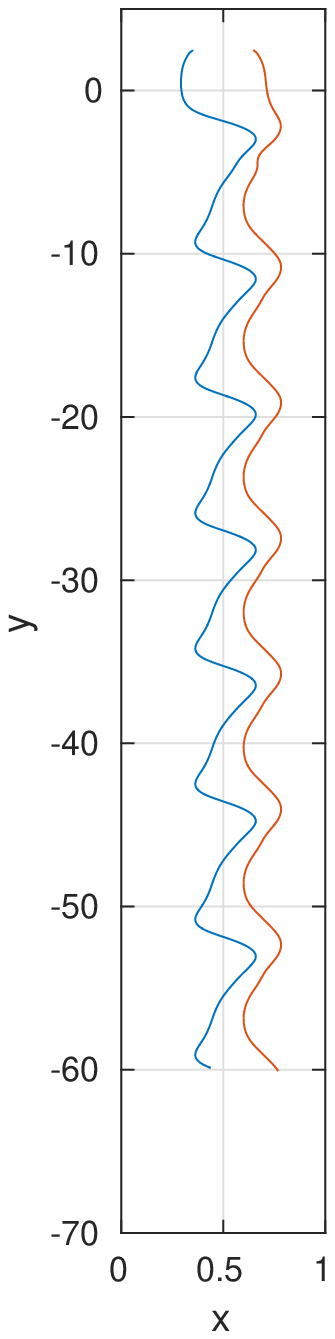}
     \end{center}  
     \vskip -2ex
     \caption{Positions of two disks   drafting, kissing, and not-chaining  for $\rho_s=1.005$ and  E=0.4  (the other associate  numbers are Re=1.0824, M=0.6846 and De=0.4330).}    \label{fig2h}
\end{figure}     
 
  \begin{figure}[tp]
     \begin{center} 
     \includegraphics[width=0.32\textwidth]{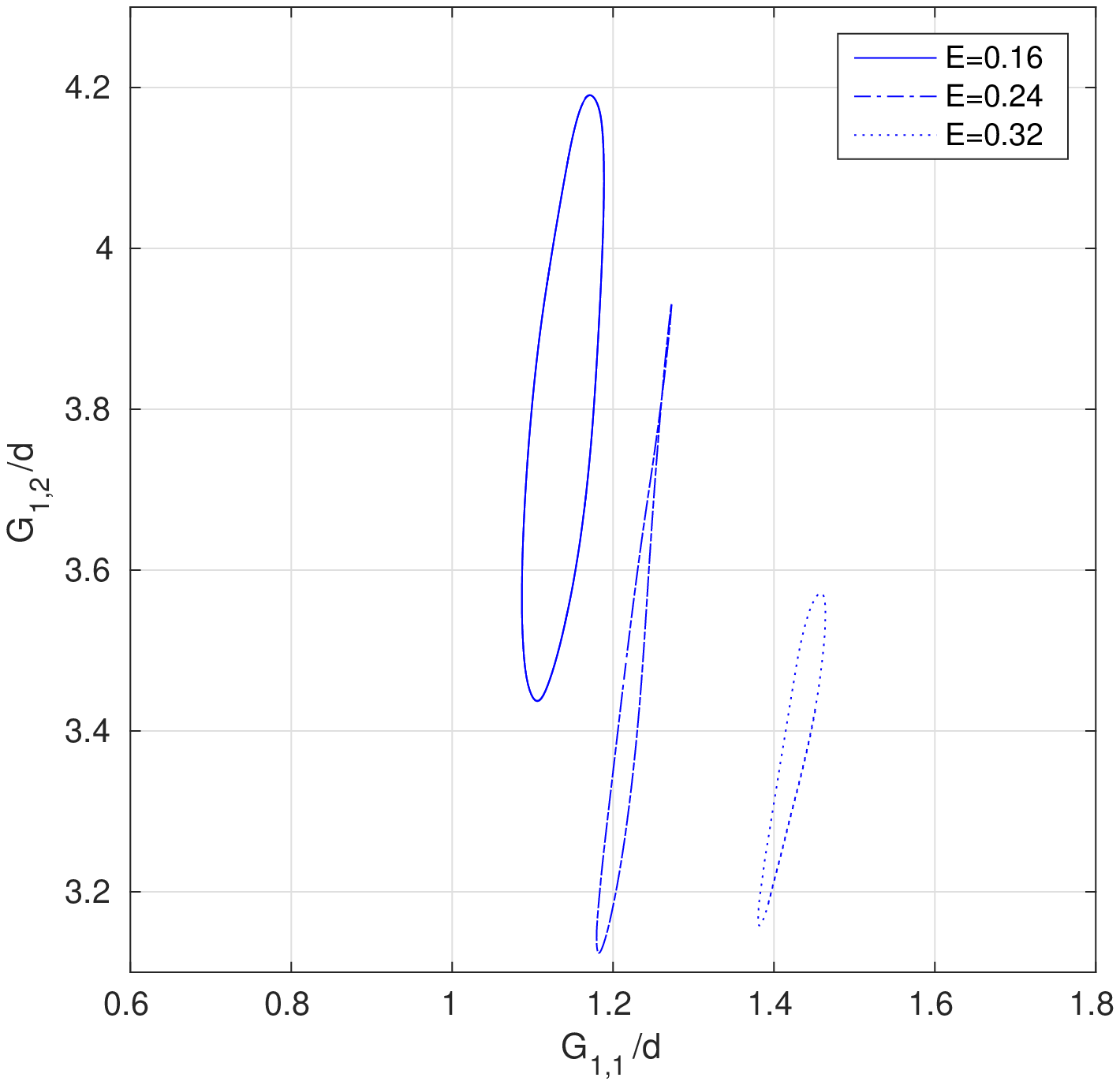}    \hskip 10pt 
     \includegraphics[width=0.32\textwidth]{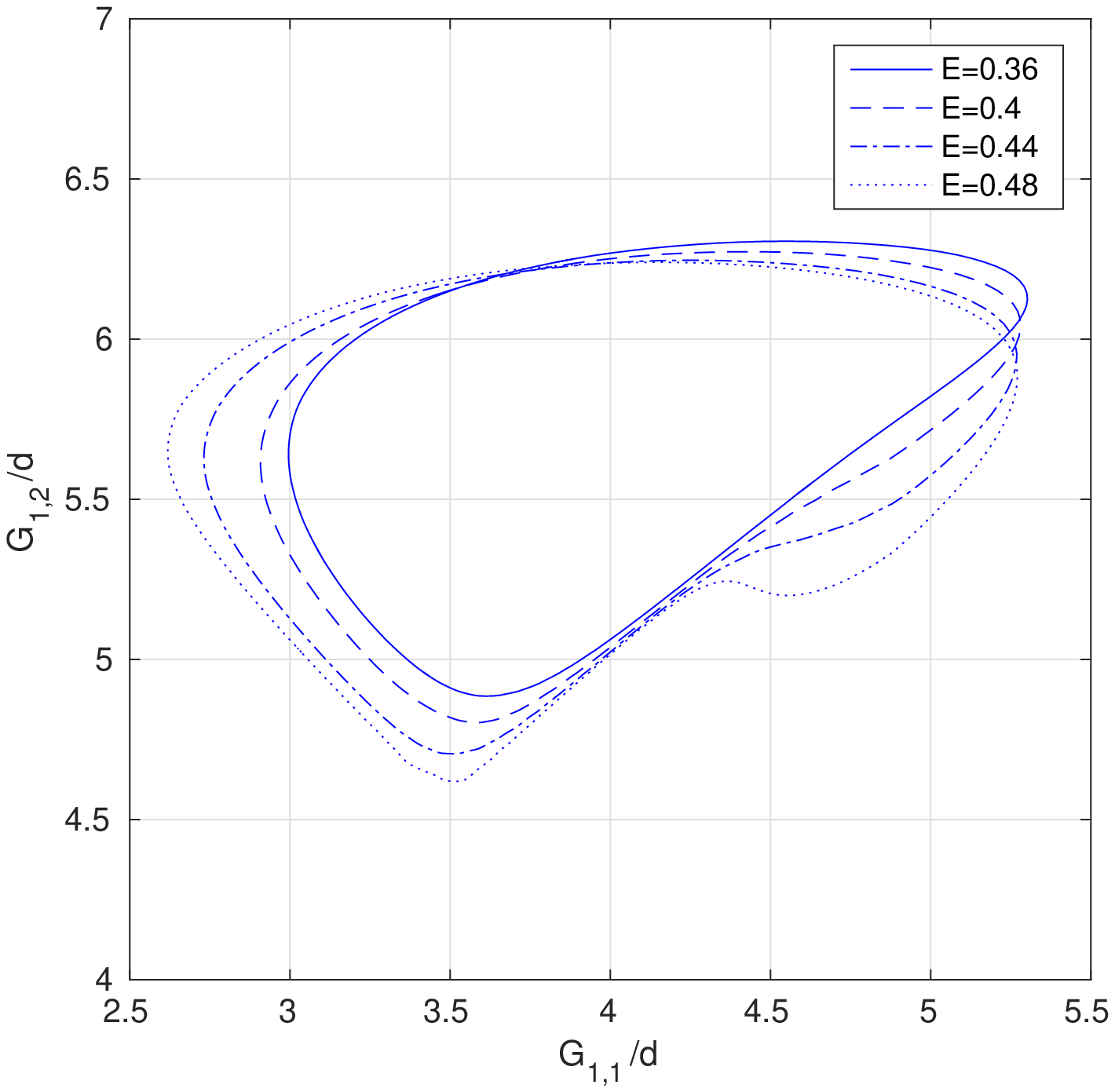}
    \end{center}  
     \caption{Trajectories of disks in phase space of the normalized distances  between the disk mass center and the left wall 
                   for $\rho_s=1.005$.  The periods of two disks interacting apart  are 21.65, 18.05 and 16.28 for  E=0.16, 0.24 and 0.32, 
                   respectively (left) and  the periods of two disks drafting, kissing, and not-chaining are  34.2, 38.45, 46.45 and 64  for E=0.36, 0.4, 0.44 and  0.48, respectively (right).}    \label{fig2g}
\end{figure}

 \begin{figure}[!tp]
     \begin{center} 
     \includegraphics[width=0.125\textwidth]{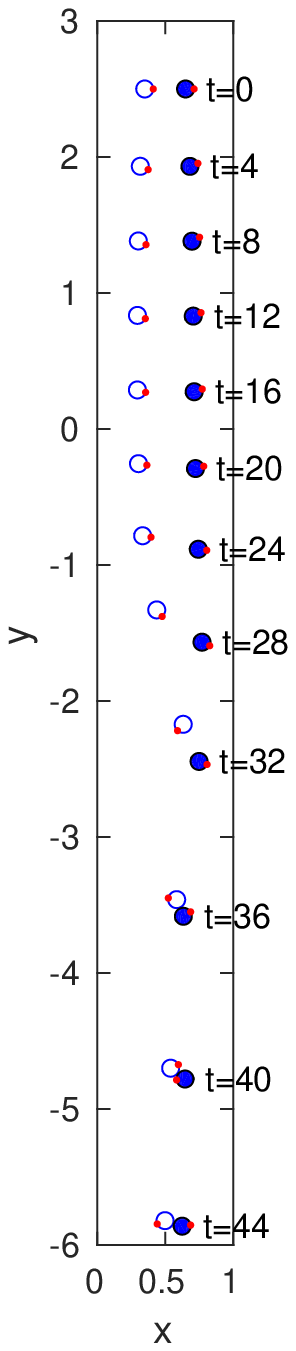}
     \includegraphics[width=0.125\textwidth]{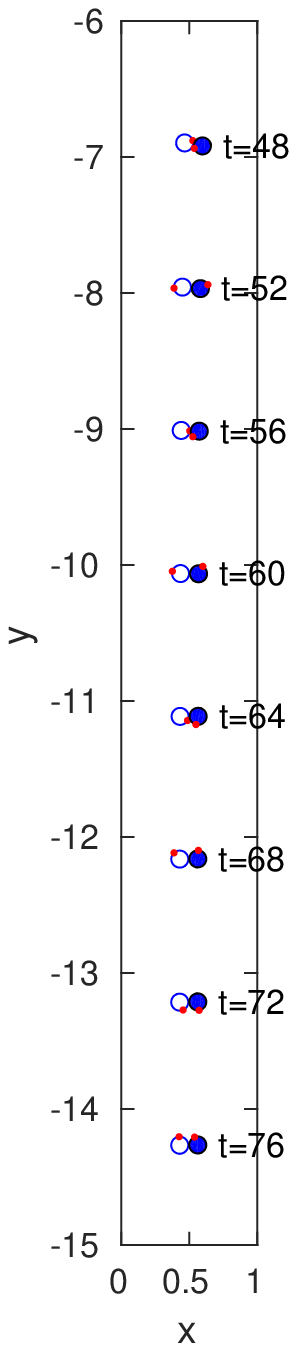}
     \includegraphics[width=0.125\textwidth]{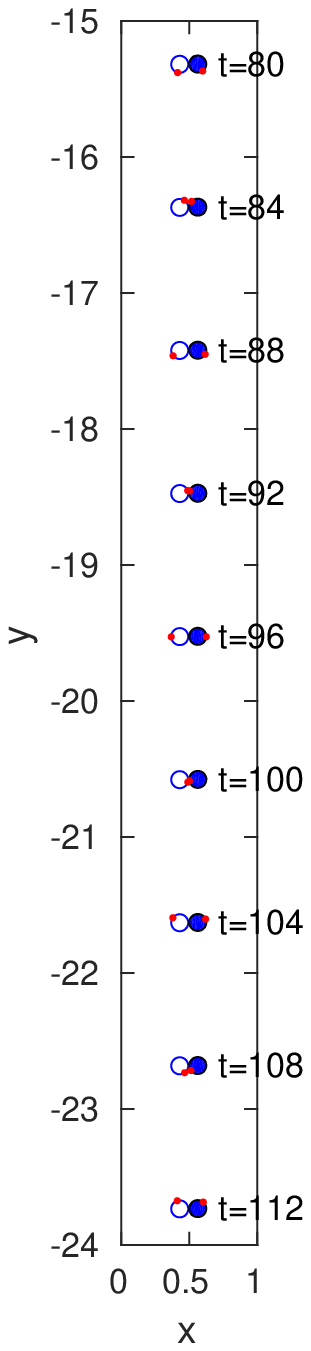}
     \includegraphics[width=0.125\textwidth]{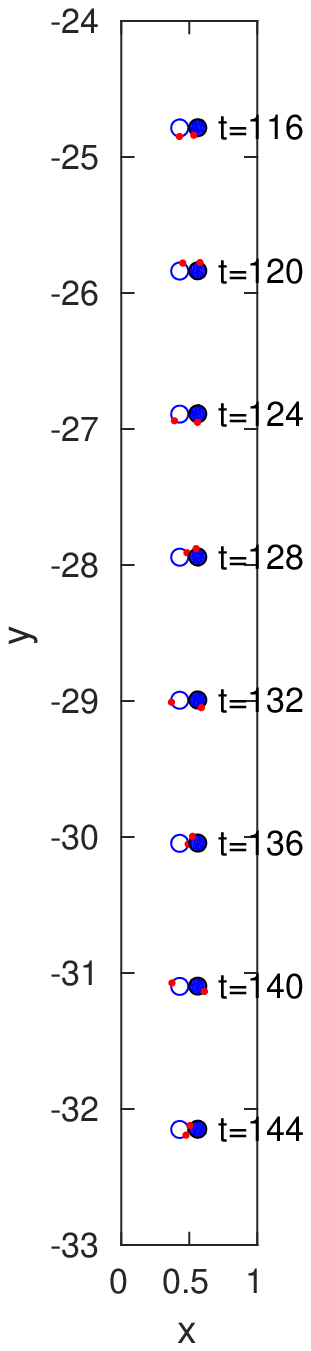}     
     \includegraphics[width=0.135\textwidth]{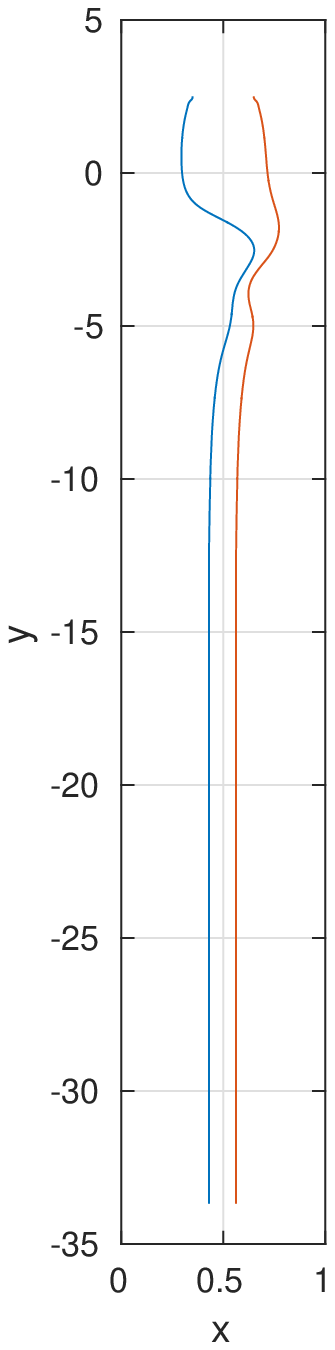}
     \end{center}  
     \vskip -2ex
     \caption{Positions of two disks  for $\rho_s=1.005$ and  E=0.56  (the other associate  numbers are Re=1.3150, M=0.9841 and De=0.7364).}    \label{fig2i}
\end{figure}

For the particles of the density $\rho_s$=1.0015, similar particle motions are obtained.  For $E  \le 0.2182$, two disks  stay separated and interact periodically. 
This periodic motion between two disks just likes the one in Fig.  \ref{fig2a} and its associated limit cycle is similar to those in the left plot in Fig. \ref{fig2d}.
For  E between 0.2424 and 1.0667, the orientation of the  disk chain oscillates first and then turns into vertical direction after the oscillation damps out 
(e.g., see  the trajectories of two disks for E=0.2424 and 0.4848 shown  in Fig. \ref{fig2f}). For E between  1.2606 and 1.6, two disks form a chain and then turn its orientation into
the falling direction right away. The tilted chain is not obtained for the values of the elasticity number considered in the phase diagram presented in Fig. \ref{fig2e}. 
The critical value of the elasticity number for having a vertical chain is somewhere in the interval  [0.2182, 0.2424].  

For the particles of slightly heavier densities $\rho_s$=  1.0035 and 1.005, we have also obtained similar kinds of particle motion for various values 
of the elasticity number as  shown together with $\rho_s$=1.0015 and 1.0025  in a phase plane in Fig. \ref{fig2e}.  But the range of the elasticity number  
for having the tilted chain is wider. Also for these relatively heavier disks, besides the typical periodical motion discussed in the above cases, there is another one 
which we call ``drafting, kissing and not-chaining''  (see Fig. \ref{fig2h}).  The limit cycles of those two types of periodic motion for $\rho_s$=1.005 are 
shown in Fig.  \ref{fig2g}. The limit cycles in the left plot in  Fig.  \ref{fig2g} are associated with the motion like the one in Fig.  \ref{fig2a}  and those in the right plot  
are associated with the drafting, kissing and not-chaining. The particle positions and  trajectories for $\rho_s$=1.005 and E=0.4  shown in Fig.  \ref{fig2h}  tell us
that every time  a chain is about to be formed after drafting and kissing between two disks,  the ``long body'' of two disks turns and then two disks break away.  
We believe that this  not being able to form a chain between two disks is due to the weak elastic force since, for  E=0.56,  an  almost horizontal and stable chain  is formed  (see Fig. \ref{fig2i}). By comparing the particle position shown in Figs.  \ref{fig2h}  and \ref{fig2i}, we observe that  two particles kiss between $t=40$ and $48$ for E=0.4 and 0.56. 
Then the pair in  Fig.  \ref{fig2h} breaks up at $t=52$ for the case of  E=0.4, but the other pair for E=0.56  remain chained.

\begin{figure}
     \begin{center} 
     \includegraphics[width=0.125\textwidth]{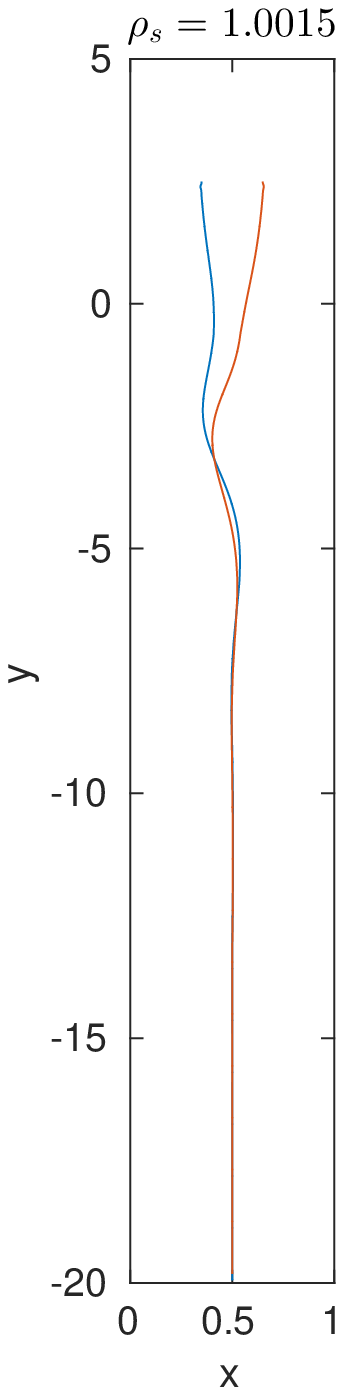}
     \includegraphics[width=0.125\textwidth]{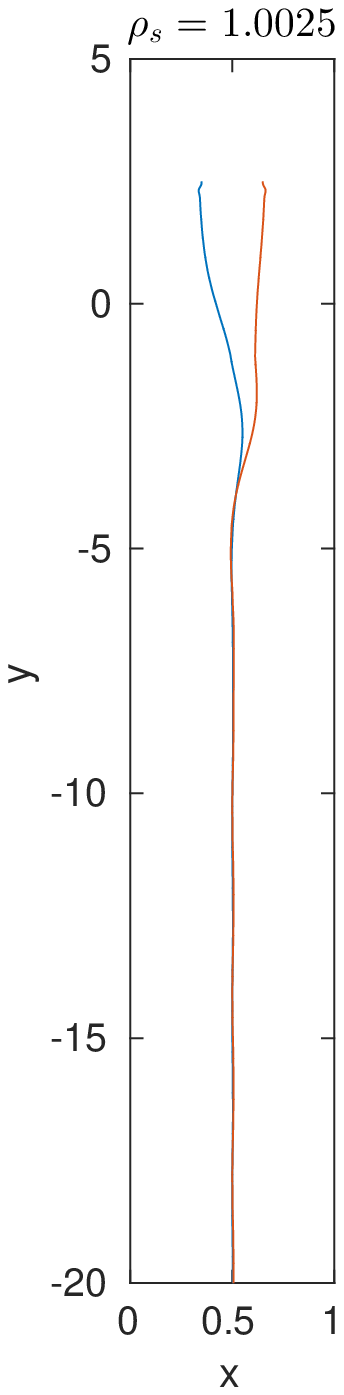}
     \includegraphics[width=0.125\textwidth]{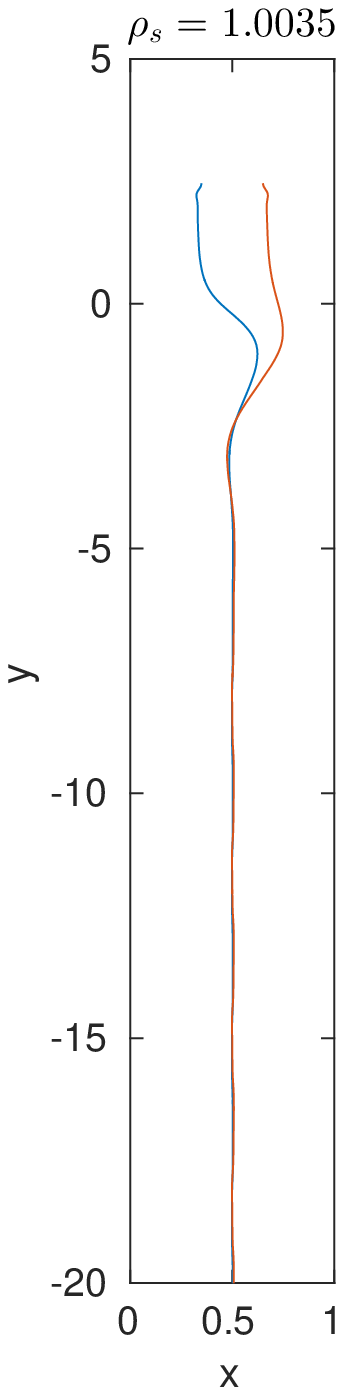}
     \includegraphics[width=0.125\textwidth]{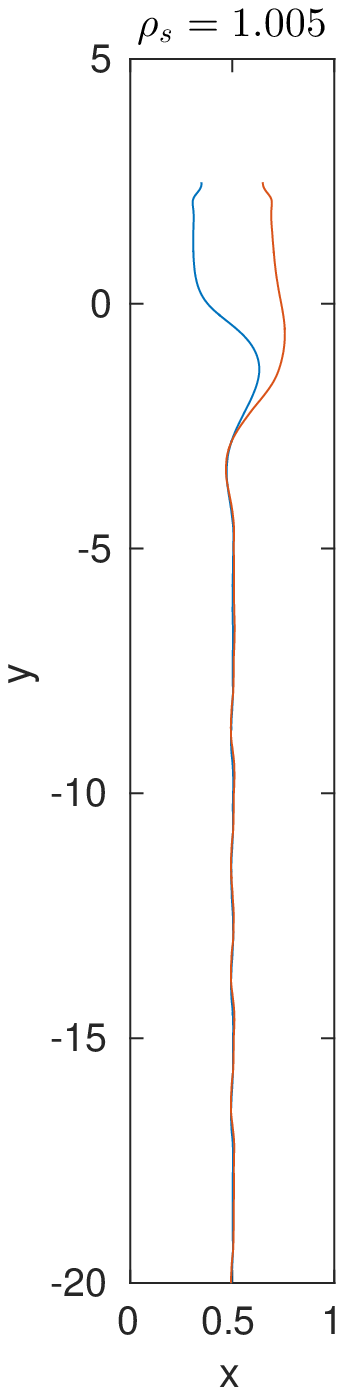}     
     \end{center}  
     \vskip -2ex
     \caption{Trajectories of two disks having vertical chain  for  four different densities  and  E=1.6. }    \label{fig2j}
\end{figure}

\begin{figure}
     \begin{center} 
     \includegraphics[width=0.125\textwidth]{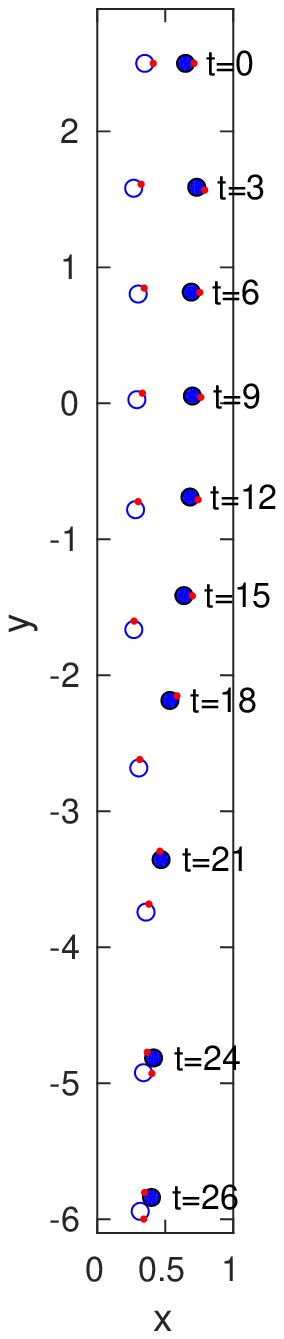}
     \includegraphics[width=0.125\textwidth]{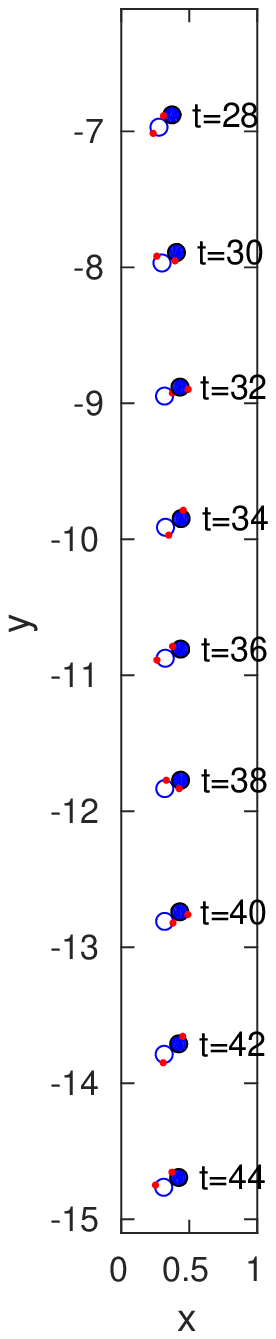}
     \includegraphics[width=0.125\textwidth]{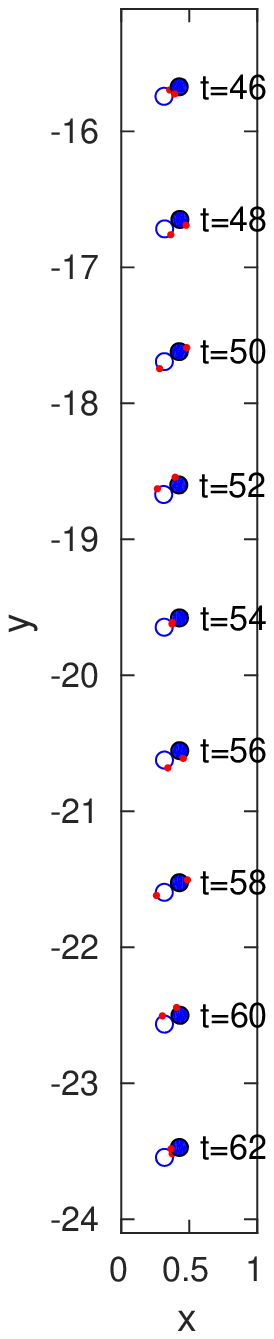}
     \includegraphics[width=0.13\textwidth]{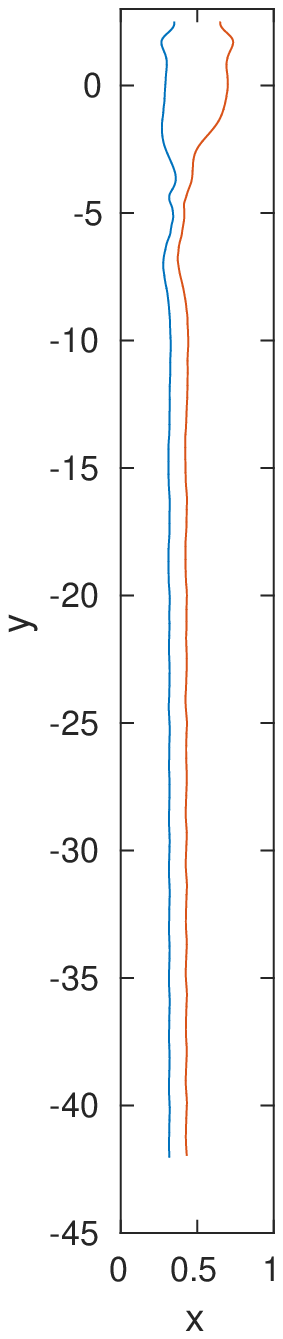}       
     \end{center}  
     \vskip -2ex
     \caption{Positions of two disks (left three) and trajectories of two disks  (right) for$\rho_s=1.01$ and  E=1.6 (the other associate  numbers are Re=2.4337, M=3.0784 and De=3.8940).}    \label{fig2k}
\end{figure}

All values of the Mach number associated with the different values of the particle density $\rho_s$ and elasticity number  are presented in  Fig. \ref{fig2e}. For each fixed value of the 
elasticity number E,  when  the particle is heavier, the Mach number is increased. For example, at E=1.6,  two disks form a chain and then its orientation turns into the falling 
direction right away for all four particle densities (see Fig. \ref{fig2j}).  The associated values of the Mach number of these four cases are 0.6697, 1.0582, 1.4004 and 1.8468
for $\rho_s=$1.0015, 1.0025, 1.0035 and  1.005, respectively.   As discussed in  \cite{huang1998}, the long body  flips falling broadside-on for Mach number greater than its 
critical value ($O(1)$).  To see the effect of the larger value of the Mach number on the chain orientation,  the numerical results of  two disks of  $\rho_s=1.01$ 
at E= 1.6  do show  that a stable tilted  particle chain is obtained  (see Fig. \ref{fig2k}). The tilt angle  of the chain is about  32 degrees
and the associated   Mach number is  M= 3.0784. For  E=1.44, similar results are also obtained. The tilt angle  of the chain is 31.59 degrees
and the associated   Mach number is  M= 2.8660. Hence the values of the elasticity number and the Mach number determine 
whether the the chain can be formed and the orientation of the chain, which is consistent with the results obtained in  \cite{huang1998} for elliptic 
particles settling Oldroyd-B fluids.


\section{Conclusion}

In this article we present a numerical study of the dynamics of two disks settling in a narrow vertical channel filled  with an Oldroyd-B fluid.
For the cases considered in this article, two  kinds of particle dynamics are obtained: (i) periodic interaction between two disks and 
(ii)  the formation of the chain of two disks.  For the periodic interaction of two disks,  two different motions are obtained: (a) two disks stay 
far apart and their  interaction is periodical, which is similar to one of the motions of two disks settling in a narrow channel filled with a Newtonian
fluid discussed in \cite{Aidun2003}  and (b)   two disks   draft, kiss and break away periodically and the chain is not formed due to the lack of strong enough 
elastic force. For the formation of  two disk chain occurred at higher values of the elasticity number, it is either a tilted chain or a vertical chain. 
The tilted chain can be obtained for either that the elasticity number  is less than the critical value for having the vertical chain or that the Mach 
number is greater than the critical value for a long body to fall broadside-on.  Hence the  values of the elasticity number and the Mach number 
determine whether the the chain can be formed and the orientation of the chain.

\section*{Acknowledgments.} 
 
We acknowledge  the support of NSF (grant DMS-1418308).

\end{document}